\documentclass[conference]{IEEEtran}
\IEEEoverridecommandlockouts

\usepackage{cite}
\usepackage{amsmath,amssymb,amsfonts, physics}
\usepackage{newtxtext,newtxmath}
\usepackage{algorithmic}
\usepackage{hyperref}
\usepackage{caption}
\usepackage{romannum}
\usepackage{float}
\usepackage{comment}
\usepackage{lipsum}
\usepackage{graphicx}
\usepackage{textcomp}
\usepackage{xcolor} 
\def\BibTeX{{\rm B\kern-.05em{\sc i\kern-.025em b}\kern-.08em
		T\kern-.1667em\lower.7ex\hbox{E}\kern-.125emX}}
\usepackage{fancyhdr}
\begin{document}
	
	\title{Design of Novel Hybrid CPDM-CO-OFDM FSO Communication System and its Performance Analysis under Diverse Weather Conditions}

	\author{\IEEEauthorblockN{Ruhin Chowdhury\textsuperscript{*,1}, A. K. M. Sharoar Jahan Choyon\textsuperscript{*,2}}
		\IEEEauthorblockA{\textsuperscript{*}\textit{Dept. of EECE, Military Institute of Science \& Technology (MIST), Dhaka, Bangladesh} \\
		\textsuperscript{1}{ruhio138@gmail.com},
		\textsuperscript{2}{choyonsharoar@gmail.com}}
		\IEEEauthorblockA{\textsuperscript{*}\textit{\textbf{Authors Contributed Equally}}}}

	\maketitle

	\begin{abstract}
	A comprehensive novel design is proposed for the free-space optical (FSO) communication system by hybridizing circular polarization division multiplexing (CPDM) with coherent optical orthogonal frequency division multiplexing (CO-OFDM) and its performance is investigated realistically under diverse turbulent weather conditions of Bangladesh. Here we consider Gamma-Gamma (G-G) distribution for the turbulent FSO channel model. Moreover, the proposed scheme presents an excellent performance since CPDM technique not only maximizes the link capacity of FSO system but also enhances the spectral efficiency (SE) of the system. Besides, multipath-fading, which is appeared during the FSO transmission, is significantly mitigated by OFDM modulation. The outcomes from simulation confirm the advantages of the proposed hybrid scheme and also it can serve as a reference for the FSO application even in the turbulent weather conditions. Performance analysis of the proposed model is described in terms of optical power spectrum (OPS), optical signal to noise ratio (OSNR), bit error rate (BER), Q factor, constellation diagrams, and eye diagrams.

	\end{abstract}


	\section{{Introduction}}
Over the last decades, FSO communication system has drawn significant attention for its outstanding advantages over optical fiber and radio frequency (RF) communications \cite{1}. FSO system is adopted for transmitting and receiving high-bandwidth digital data operating in the span from 100 Mbps to 1.2 Tbps over short distances with no optical spectrum licensing \cite{2,3}. Thus, FSO link is cost effective compared to RF communication. Besides, FSO communication system is easy to install and feasible. This optical wireless system exploits low powered LASER or LED as an optical source which is invulnerable to electromagnetic interference (EMI), jamming and detection offering covert connections, high security and easy communication over RF and microwave communications \cite{4}-\cite{JOC}. 
Since FSO is a line of sight (LOS) and point-to-point communication system across the atmosphere, the efficacy of this system depends greatly on weather conditions, atmospheric pressure, temperature, etc. It is highly vulnerable to absorption, scattering, scintillation, temperature variations, irregular refractive index, etc. Thus, weather conditions, for instance, rain, fog, snow, haze, etc limit the performance and link distance of FSO system \cite{6}. Even in clear weather conditions, turbulences affect the system performance quite a lot. Besides, there are always water molecules and gaseous particles in the air which mitigate the light beam intensity and causes multipath fading. Weather of different geological locations effects the available FSO link diversely. For instance, fog in temperate regions, heavy rainfall in tropical regions influence the FSO link greatly. Besides, haze induces  by smoke, mist and other dry particles have low impact on the optical power compare with rain resulting in Mie scattering of optical signal. This scattering will induce attenuation of optical received power, thus reduce the availability of FSO link for a given link distance \cite{Is}. As Bangladesh lies near equator and has a subtropical monsoon climate, we have considered fog and rain as our main weather conditions. Researchers investigate that, in Bangladesh, atmospheric attenuations can arouse to 23.12 dB/km and 12.47 dB/km for heavy rain and light fog environments \cite{7}. The requisite requirements to achieve the best performance of an FSO system is to mitigate multipath channel fading along with the effects of atmospheric turbulence as much as possible while keeping the capacity high. An easy solution is to create several short communication links by breaking long ones, making the FSO link more reliable and diminishing the effects of atmospheric turbulence. Furthermore, deploying a hybrid modulation scheme can also reduce turbulence effects and channel fading. Combining the benefits of various traditional modulations, the novel hybrid system can enhance the SE and the BER performance by modulating optical signals from different aspects.

Recently, the performance of FSO system is enhanced by introducing polarization division multiplexing (PDM) technique which divides the Laser light into two orthogonal states of polarization (SOP), transmitting different signals over those polarization states ultimately doubling the SE of the system as well as the capacity \cite{8}. A direct detection (DD) OFDM technique is proposed in \cite{9} focusing on BER analysis with Log and G-G model. Performance analysis of Digital Modulation CO-OFDM 16-QAM is carried out showing superior BER and SER performance with better receiver sensitivity compared to other modulation formats and intensity-modulated direct detection (IM/DD) \cite{10}. Besides, robustness against channel fading effects can also be further improved using coherent scheme\cite{10a}. A novel hybrid PDM-OFDM model is proposed using OptiSystem in \cite{11} with improved performance increasing the user capacity, SE and reducing multipath fading for FSO transmission system. In \cite{12}, a PDM-CO-OFDM FSO communication system is introduced to reduce the influence of scintillation through atmospheric FSO channel resulting in long distance propagation under strong turbulence as depolarizing property is weakest in the atmosphere. 

But as the demand increases, the channel capacity of optical communication must increase. To overcome this growing demand and to tackle the challenge of increasing channel capacity, CPDM can be used. A CPDM system is basically an integration of two PDM systems involving two orthogonal- right circular polarization (RCP) and left circular polarization (LCP), thus quadrupling the capacity and the SE of optical system \cite{RM}-\cite{choyon}.  However, this system is quite vulnerable to multipath channel fading. To solve this problem one technique is to use coherent detection along with OFDM. OFDM technique involves transmitting user data over many parallel subcarriers with low data rates using fast Fourier transform (FFT) to cancel ISI and CO detection brings linearity to the OFDM in both RF to optical up/down converter \cite{8,14}. OFDM provides coherent detection with robustness, easy phase and channel estimation, and high SE \cite{15}. Thus, the CO-OFDM system can boost receiver sensitivity and further reduce polarization mode dispersion (PMD) and chromatic dispersion (CD) \cite{16}. Integrating CPDM with CO-OFDM brings all these advantages together and reduces the effects of multipath fading while increasing channel capacity and SE for both wired and wireless communication.

Inspired by the above-mentioned investigations, a new hybrid scheme is proposed for the FSO link combining CPDM and CO-OFDM using QPSK modulation format to boost the channel capacity and expand SE without changing the bandwidth of the transmitter and reduce multipath fading. Performances are investigated realistically to understand the feasible limit range and advantages needed for the design of the FSO link under the turbulent weather conditions of Bangladesh, where the atmospheric attenuations are taken from \cite{7}. Here we consider G-G distribution for the turbulent atmospheric channel. The remaining part of this article is arranged as: Section \ref{sys} explains the detailed system design for the proposed hybrid CPDM-CO-OFDM FSO link. Section \ref{result} elaborates the results drawn from the simulation using OptiSystem 17. Finally, Section \ref{conclusion} summarizes the article with possible future research to extend this work.

	\begin{figure*}
	\centering
	\includegraphics[width=7in,height=9in,keepaspectratio]{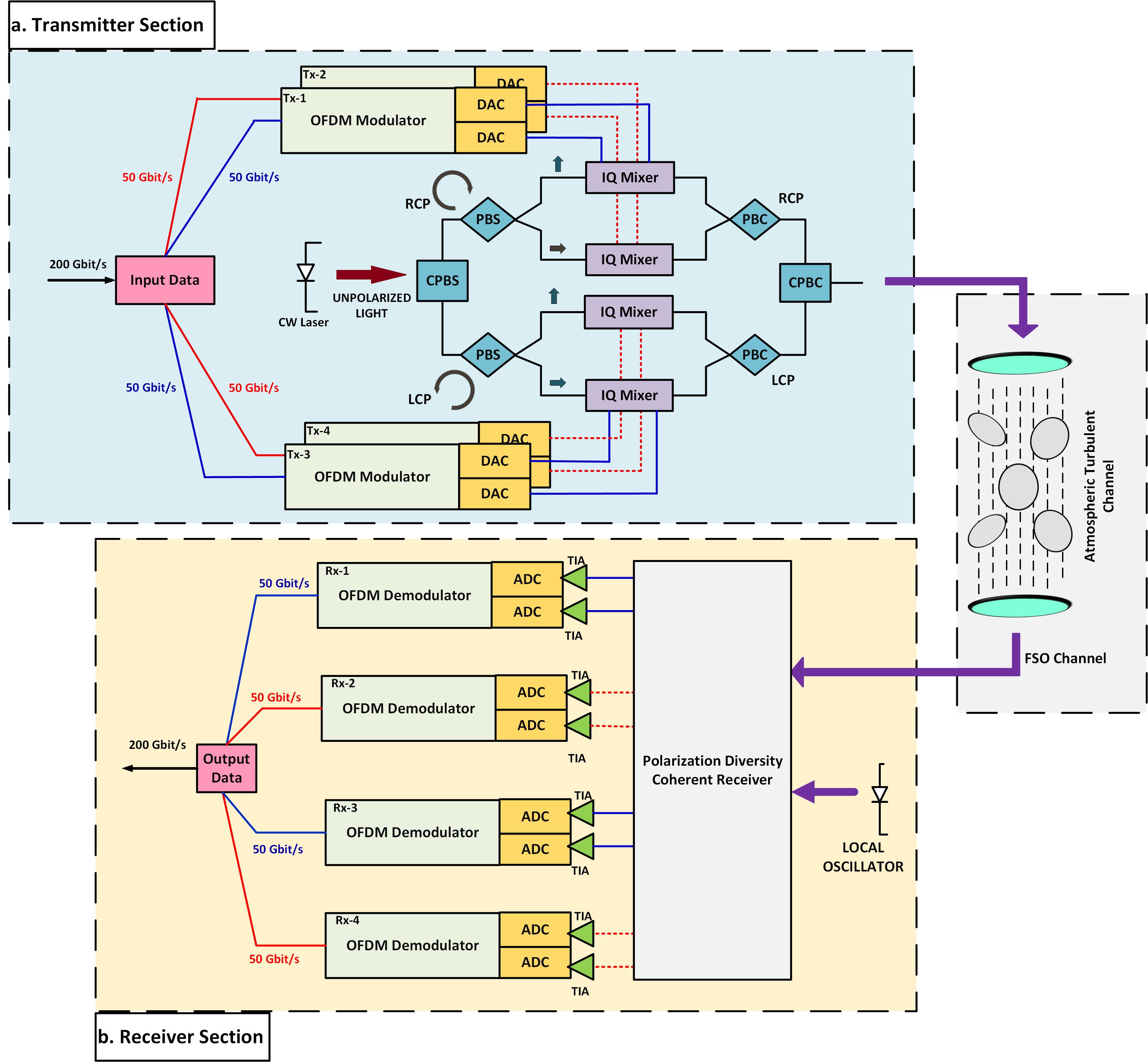} 
	\caption{\centering Design of Proposed Hybrid CPDM-CO-OFDM FSO Link. }\label{hybrid}	
    \end{figure*}

   \begin{figure*}
	\centering
	\includegraphics[width=5in,height=3in,keepaspectratio]{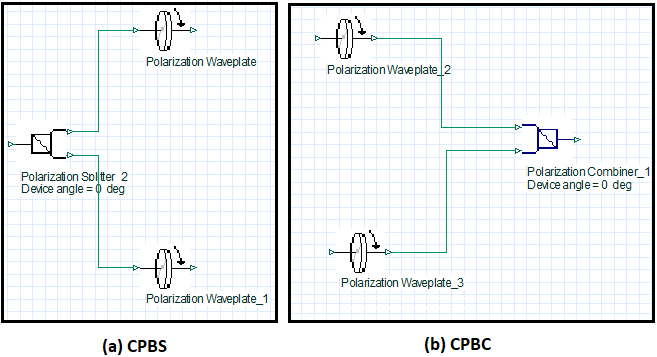} 
	\caption{\centering Design of (a) CPBS (b) CPBC.}\label{CPBS-C}	
    \end{figure*}

	\begin{figure*}
	\centering 
	\includegraphics[width=7in,height=9in,keepaspectratio]{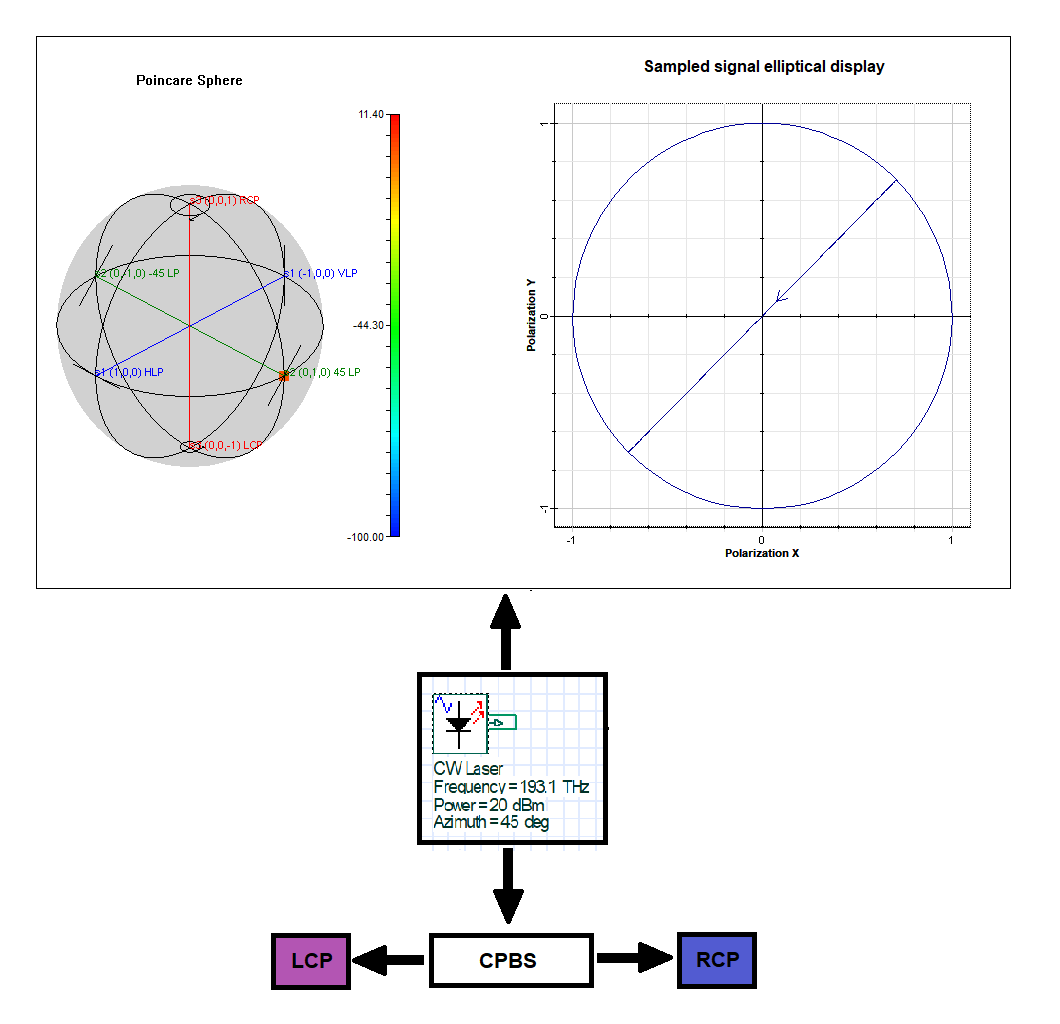} 
	\caption{\centering SOP of CW laser.}\label{PCW}	
    \end{figure*}

	\begin{figure*}
	\centering 
	\includegraphics[width=7in,height=9in,keepaspectratio]{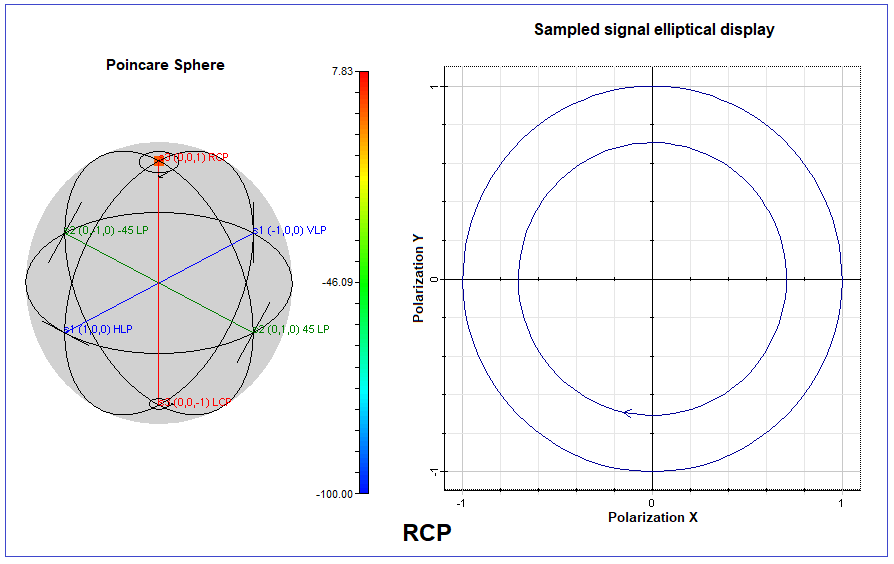} 
	\caption{\centering SOP of optical carrier signal before IQ mixer (RCP).}\label{PRCP}	
    \end{figure*}
    
    \begin{figure*}
	\centering 
	\includegraphics[width=7in,height=9in,keepaspectratio]{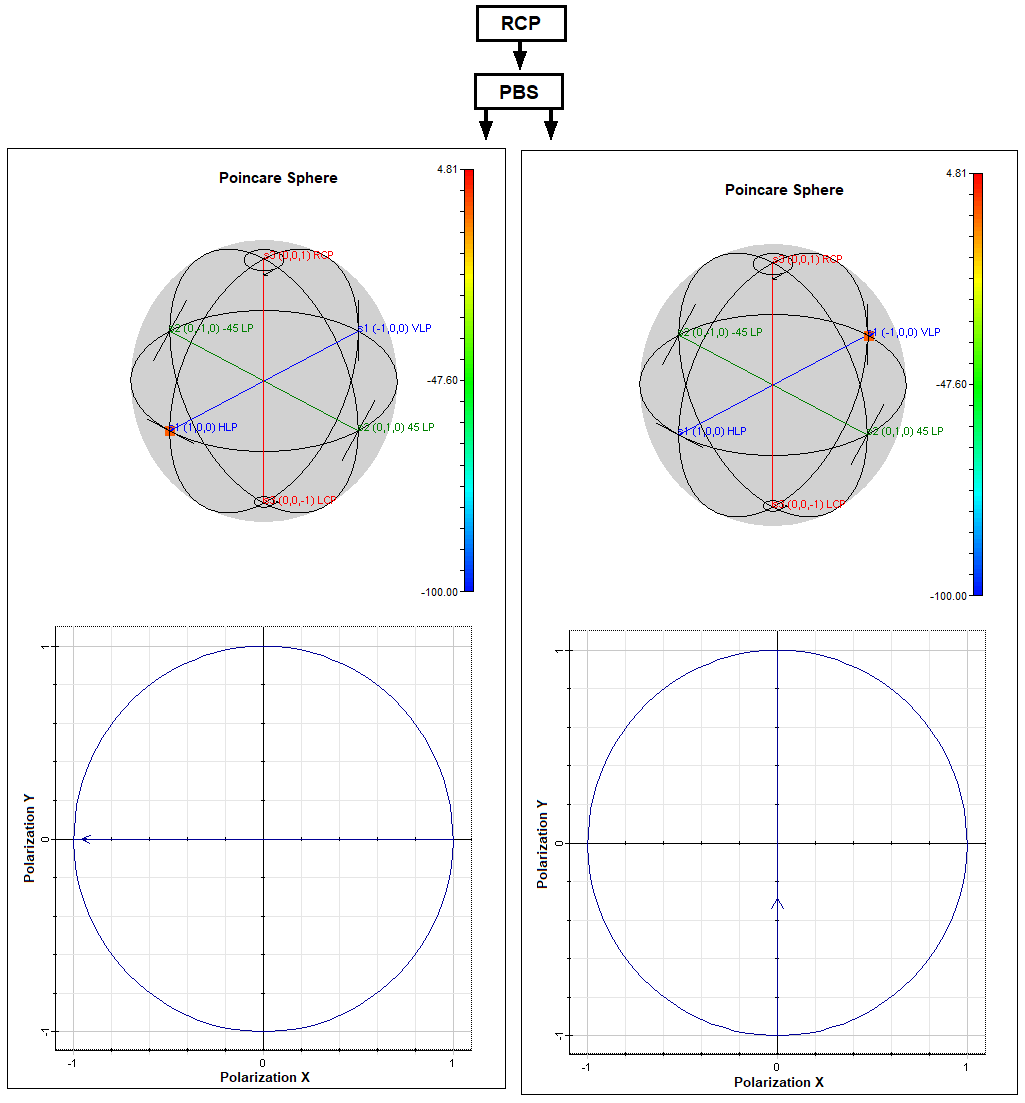} 
	\caption{\centering SOP of RCP's linear component-HLP \& VLP before IQ mixer.}\label{PRCP1}	
    \end{figure*}
    
	\begin{figure*}
	\centering 
	\includegraphics[width=7in,height=9in,keepaspectratio]{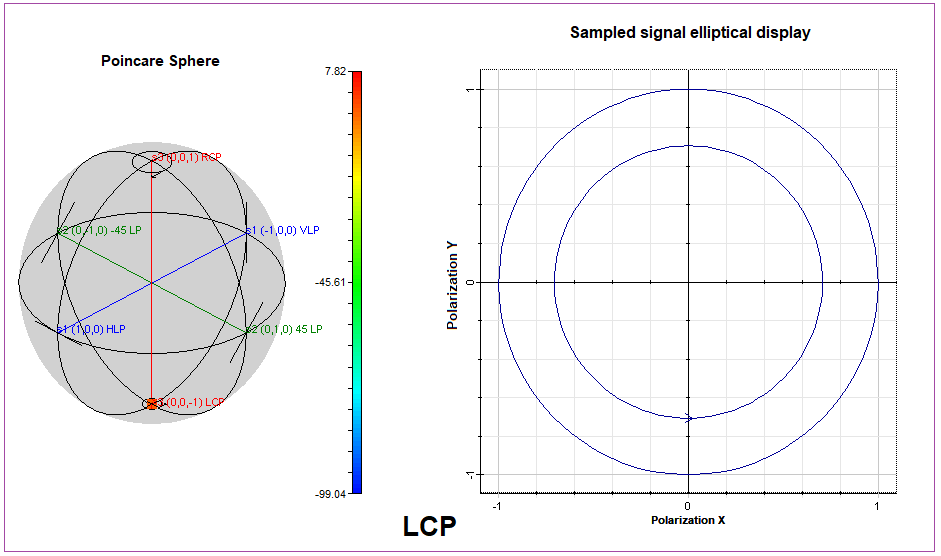} 
	\caption{\centering SOP of optical carrier signal before IQ mixer (LCP).}\label{PLCP}	
    \end{figure*}
    
    \begin{figure*}
	\centering 
	\includegraphics[width=7in,height=9in,keepaspectratio]{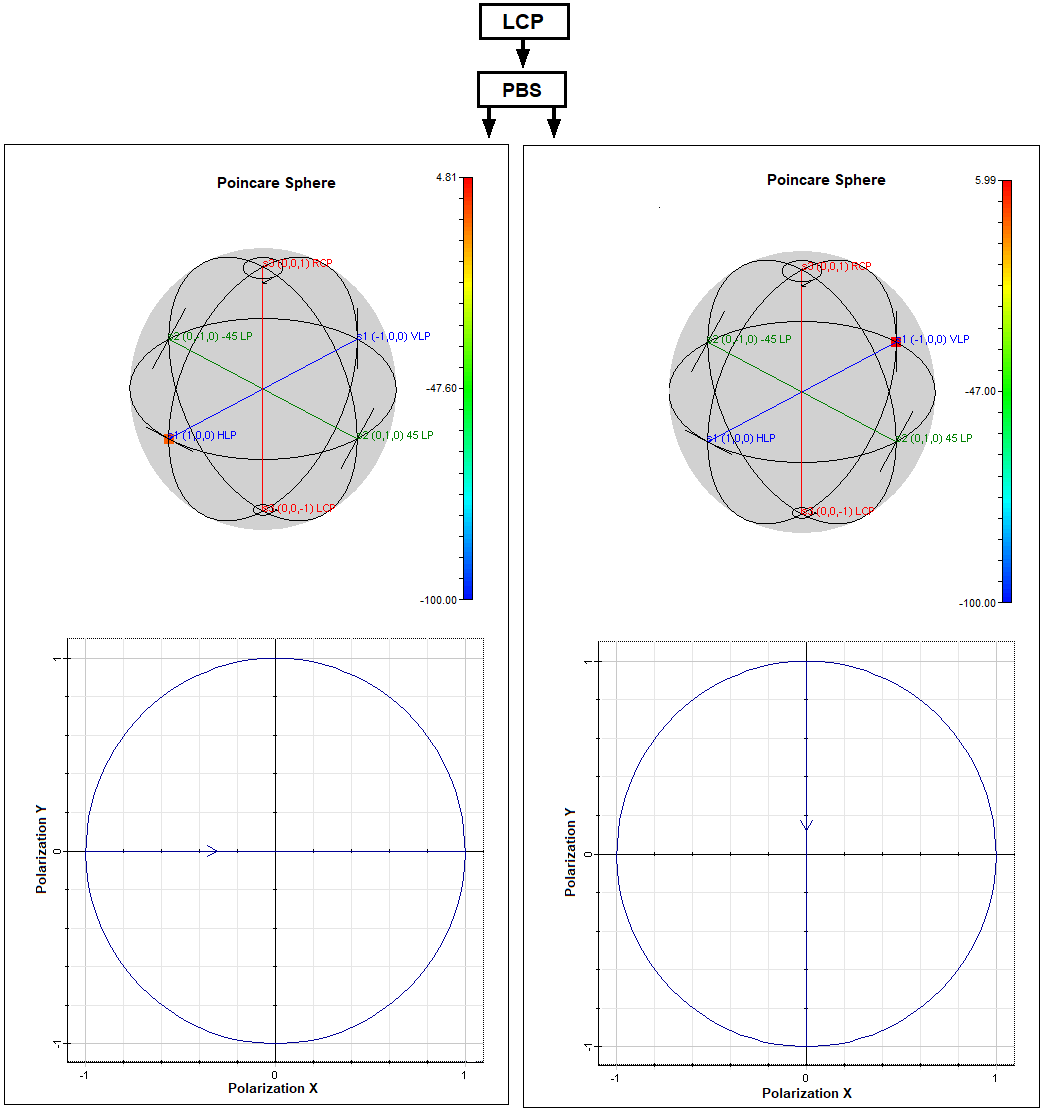} 
	\caption{\centering SOP of LCP's linear component-HLP \& VLP before IQ mixer.}\label{PLCP1}	
    \end{figure*}
    
	\begin{figure*}
	\centering 
	\includegraphics[width=7in,height=9in,keepaspectratio]{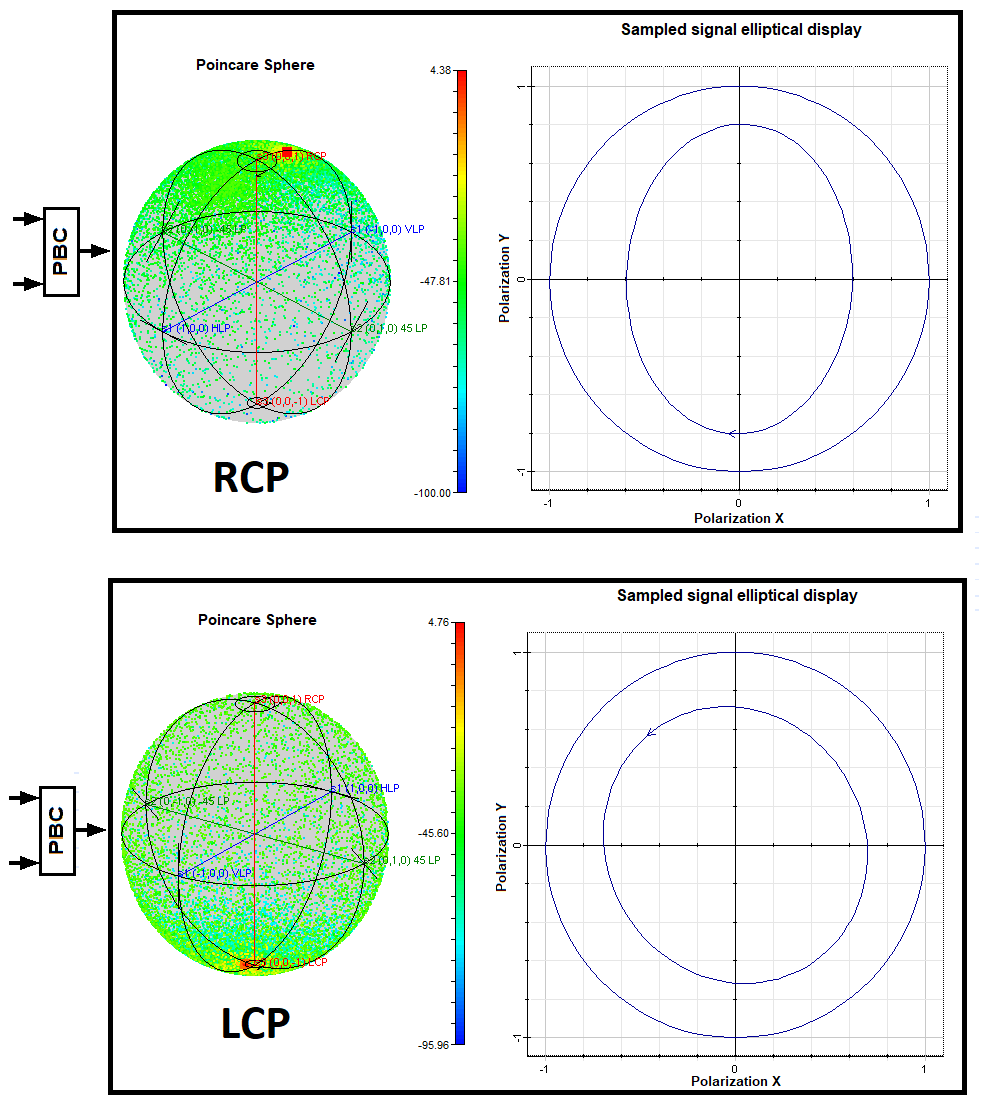} 
	\caption{\centering SOP of modulated optical signal after IQ mixer (RCP \& LCP).}\label{pol-af}	
    \end{figure*}
    
	\begin{figure*}
	\centering 
	\includegraphics[width=7in,height=9in,keepaspectratio]{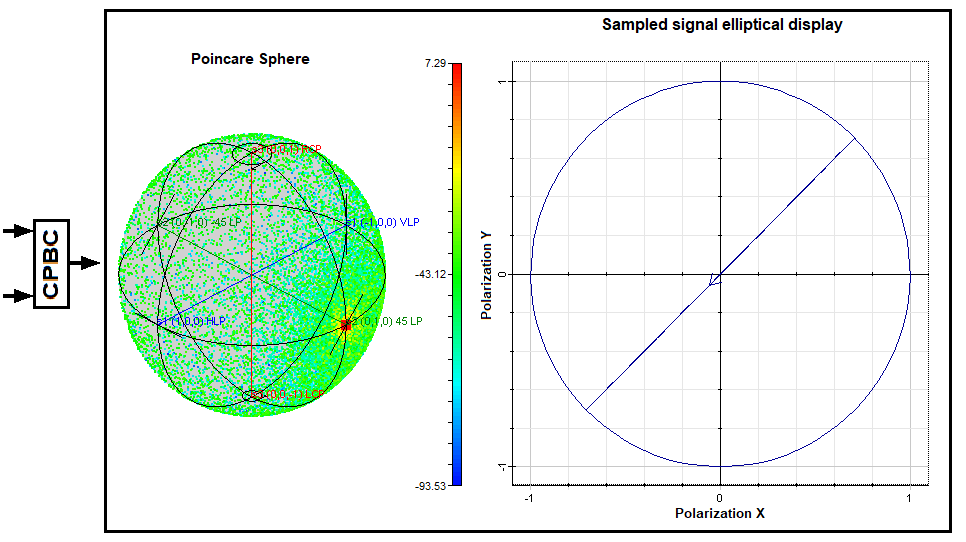} 
	\caption{\centering SOP of modulated circular-polarized (combination of RCP \& LCP) optical signal after IQ mixer.}\label{pol-af1}	
    \end{figure*}

	\begin{figure*}
	\centering 
	\includegraphics[width=7in,height=9in,keepaspectratio]{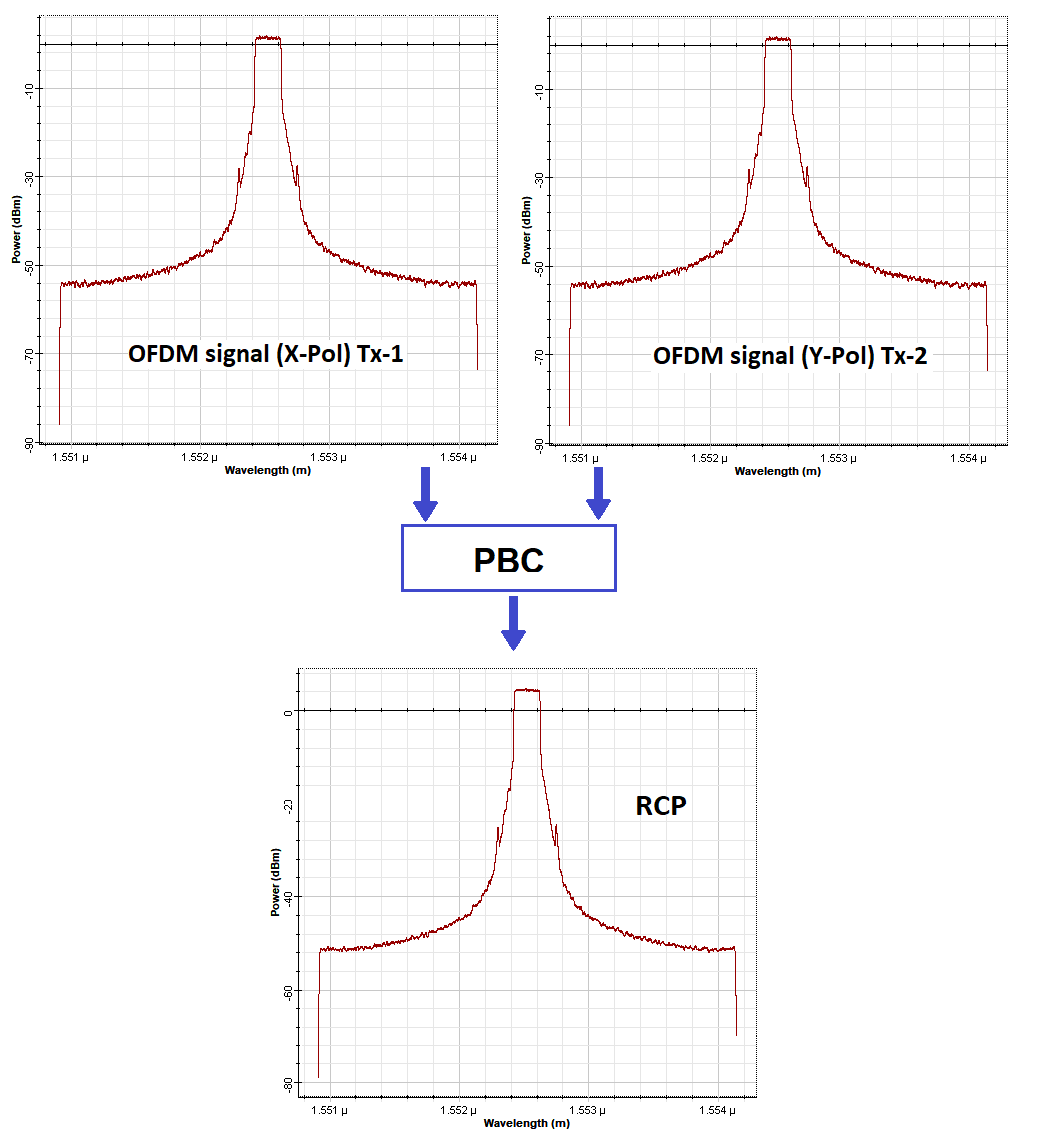} 
	\caption{\centering OPS of OFDM modulated RCP signal.}\label{OPS1}	
    \end{figure*}
    
	\begin{figure*}
	\centering 
	\includegraphics[width=7in,height=9in,keepaspectratio]{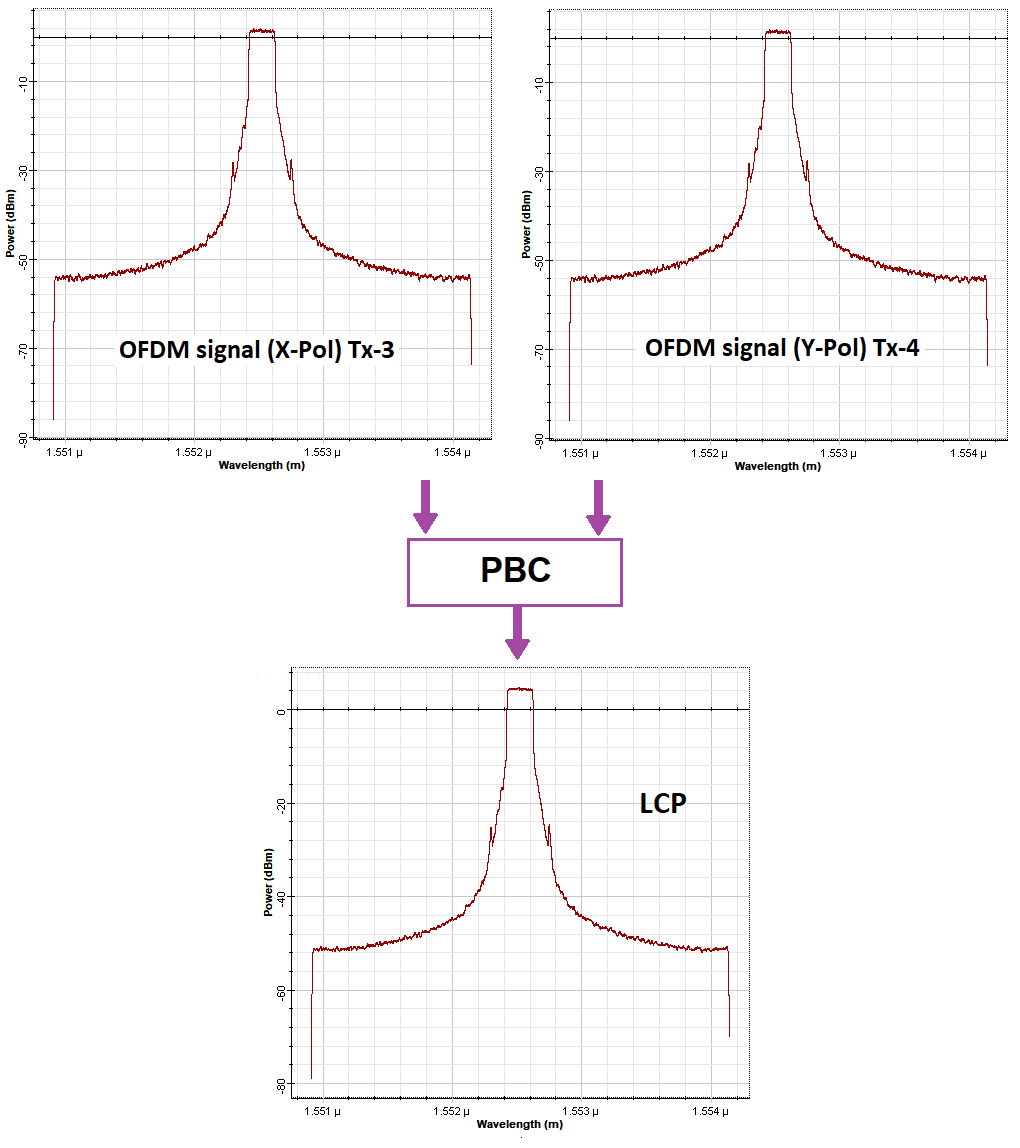} 
	\caption{\centering OPS of OFDM modulated LCP signal.}\label{OPS2}	
    \end{figure*}

    \begin{figure*}
	\centering 
	\includegraphics[width=7in,height=9in,keepaspectratio]{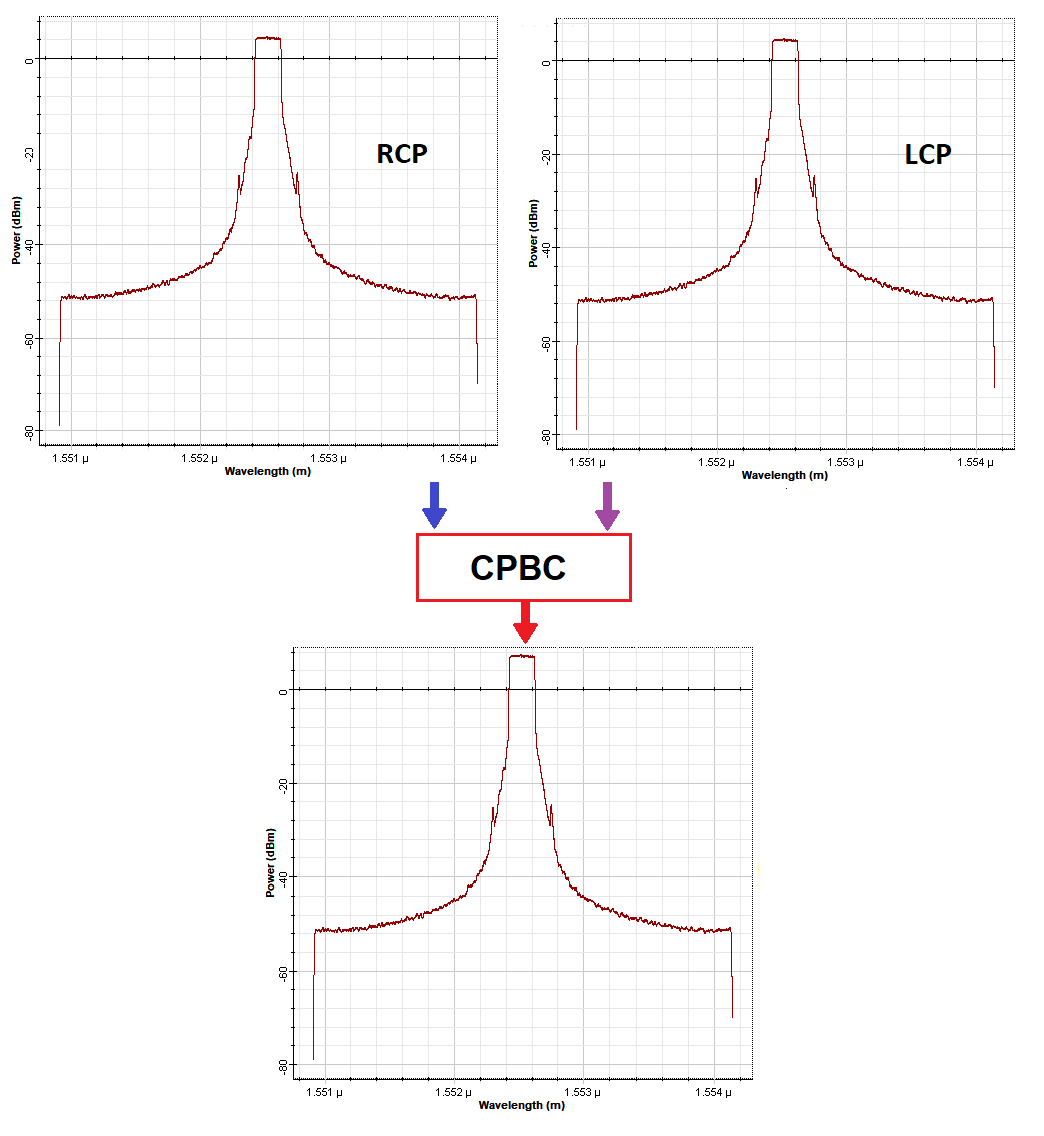} 
	\caption{\centering OPS of OFDM modulated signal (combination of RCP \& LCP).}\label{OPS3}	
    \end{figure*}

    \begin{figure*}
	\centering
	\includegraphics[width=7in,height=9in,keepaspectratio]{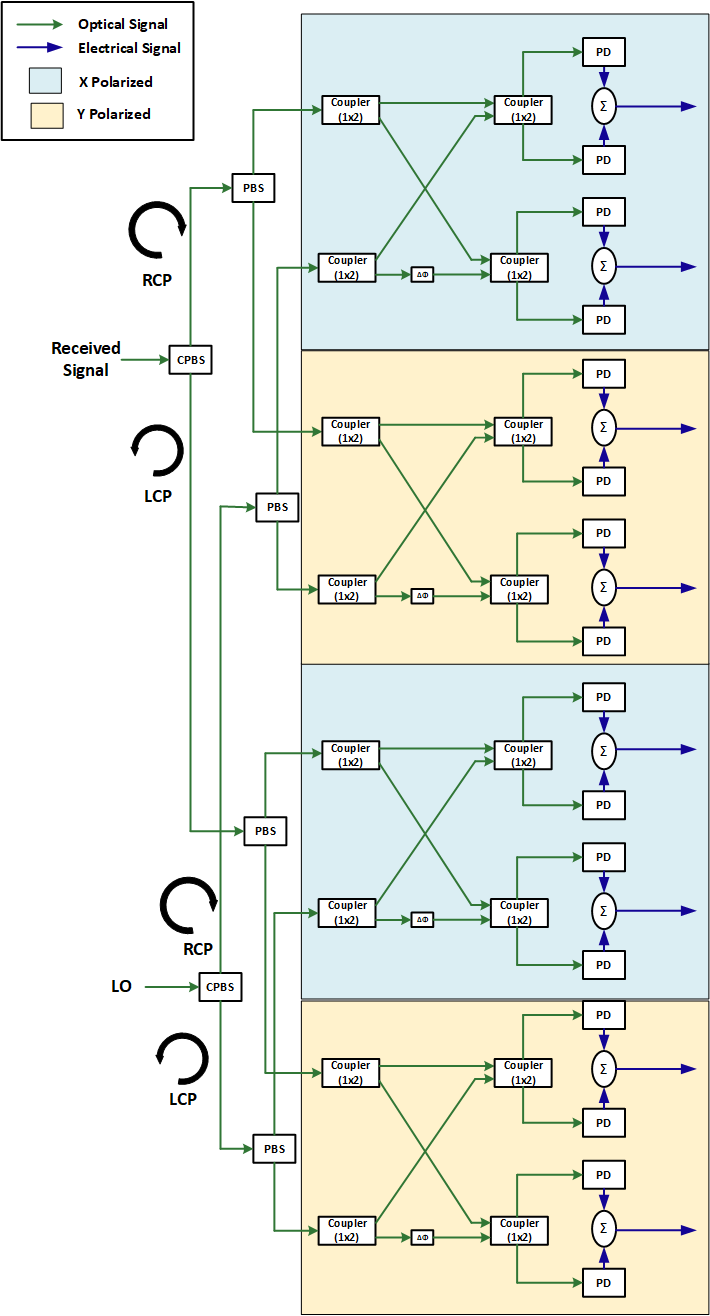} 
	\caption{\centering Design of Polarization Diversity Coherent Optical Receiver.}\label{Rx-h}	
    \end{figure*}
    
    \begin{figure*}
	\centering
	\includegraphics[width=7in,height=4in,keepaspectratio]{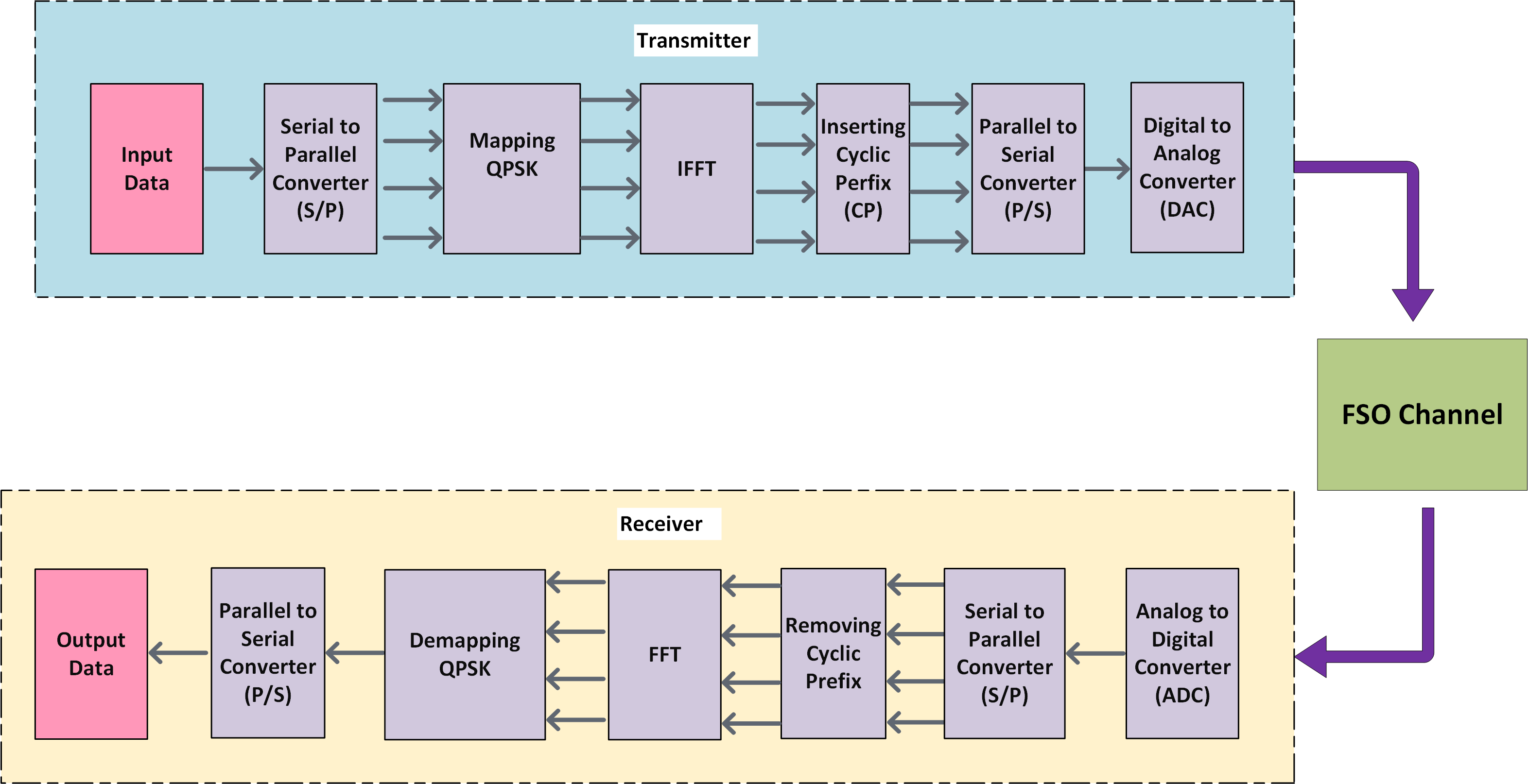} 
	\caption{\centering Block diagram of OFDM transmitter and receiver. }\label{OFDM}	
    \end{figure*}

\newpage
	
	\section{\textbf{System Design and FSO Channel Modeling}}\label{sys}

	\subsection{\textbf{Novel System Design for Proposed Hybrid CPDM-CO-OFDM FSO Link}}
		In this section, the novel hybrid CPDM-CO-OFDM FSO system model, illustrated in Fig \ref{hybrid}, is designed and simulated by OptiSystem 17 software. The transmitter section comprises of a CW Laser diode as source, a circular polarization beam splitter (CPBS), a circular polarization beam combiner (CPBC), two polarization beam splitter (PBS), two polarization beam combiner (PBC), four optical IQ mixers and four OFDM modulators. The CW Laser power is taken as 20 dBm, frequency is set to 193.1 THz and is operated in 45 degrees azimuth. Fig \ref{CPBS-C} represents the design of CPBS and CPBC. A CPBS is a combination of a PBS following by two quarter waveplates and a CPBC is a combination of two quarter waveplates following by a PBC. The input laser power at the transmitter section is divided into two circularly polarization states-Right Circular and Left Circular- using a CPBS. The outputs of a CPBS is fed into two PBS. Each PBS divides the input circularly polarized laser power into two orthogonal polarization states (Horizontal \& Linear) and is fed into optical IQ Mixer which is used as carrier. A CPBS can split an unpolarized beam of light into another set of orthogonal pair having circular polarization as Right Circularly Polarized (RCP) and Left Circularly Polarized (LCP) beams, described in the references \cite{RM}-\cite{choyon}. Again, a PBS is used for both of RCP and LCP to get 2 sets of linearly polarized light. Thus, we can achieve 4 independent channels. A CPDM is a combination of two independent PDM systems, which are independent channels, thus the output of two PBSs (4 independent channels) must be independent, explicated in the references \cite{RM}-\cite{choyon}. Moreover, the Poincare Sphere with elliptical display, depicted in Fig \ref{PCW} - Fig \ref{PLCP1}, also proves this independency. Hence, the system capacity as well as the spectral efficiency of a CPDM system is doubled than an existing PDM system. The Shannon capacity limit (SCL) is defined as the theoretical maximum amount of information that can be achieved during the transmission of the data between the Tx and Rx and the SCL is a useful metric for determining the spectral efficiency of the system \cite{shanon2}-\cite{shanon}. The Shannon capacity $C$ (unit: b/s) and spectral efficiency $S$ (unit: b/s/Hz) are given by \cite{shanon}-\cite{shanon1}:
		
		\begin{equation}	
	           C= m B \log_{2}(1+SNR)  
	    \end{equation}
	    \begin{equation}	
	           S= m \log_{2}(1+SNR)   
	    \end{equation}
	   
	    Where, $m$ represents the polarization factor; for single polarization, $m=1$; for PDM, $m=2$; for CPDM, $m=4$; and $B$ is the electrical bandwidth of the modulated signal and $SNR$ is the average electrical signal-to-noise ratio.

		Meanwhile, to generate 200 Gbps information signal, a pseudo random bit sequence generator (PRBS) is employed and divided into four equal segments each transmitting 50 Gbps data. Each of the four segments is fed into an OFDM modulator resulting in a real and imaginary part followed by a digital to analog converter (DAC) and finally converted into an analog signal. To subdue aliasing components, a low pass filter (LPF) is necessary at the output of a DAC. As filters are used for every DAC and analog to digital converter (ADC), they are not presented in Fig \ref{hybrid} and Fig \ref{OFDM} for simplicity. After filtering, the in-phase (I) and quadrature-phase (Q) signals are fed to the optical IQ mixer. A complex Mach–Zehnder modulator (CMZM) is basically an IQ mixer consisting of two parallel LiNb MZMs with a relative phase shift of 90 degrees between them \cite{17}. Therefore, each PBS generates two modulated data (X \& Y Polarized). In this proposed model, as we use two PBS, we get two sets of X \& Y Polarized modulated data. Each set of X \& Y Polarized modulated data are combined using a PBC. The outputs of the two PBCs are RCP and LCP respectively which are converted into linear polarized light by a CPBC and sent through the FSO channel. The states of polarization (SOP) of optical carrier signal before modulation and modulated optical signals after going through IQ mixers are expounded in detail with Poincare sphere by Fig \ref{PCW}, Fig \ref{PRCP}, Fig \ref{PRCP1}, Fig \ref{PLCP}, Fig \ref{PLCP1}, Fig \ref{pol-af} and Fig \ref{pol-af1}, respectively. Besides, Fig \ref{OPS1}, Fig \ref{OPS2} and Fig \ref{OPS3} elaborate the OPS of OFDM modulated signals after IQ mixers, PBCs and CPBC. To resolve power attenuation, pre-optical amplifier of gain 15 dB is used. As a result, 50 Gbps signal is applied to each OFDM modulator and the achieved signal rate of the CPDM transmitter becomes 200 Gbps using four IQ mixers whereas, the signal rate of a PDM transmitter would have been 100 Gbps using only two IQ mixers. The SE of a CPDM transmitter is thus doubled comparing to a PDM transmitter without altering the bandwidth of the transmitter as explained. The design parameters of the hybrid CPDM-CO-OFDM system are listed in Table \ref{para} while Table \ref{attenuation} illustrates the attenuation parameters for various weather conditions of Bangladesh.

    \begin{figure*}
	\centering
	\includegraphics[width=7in,height=5in,keepaspectratio]{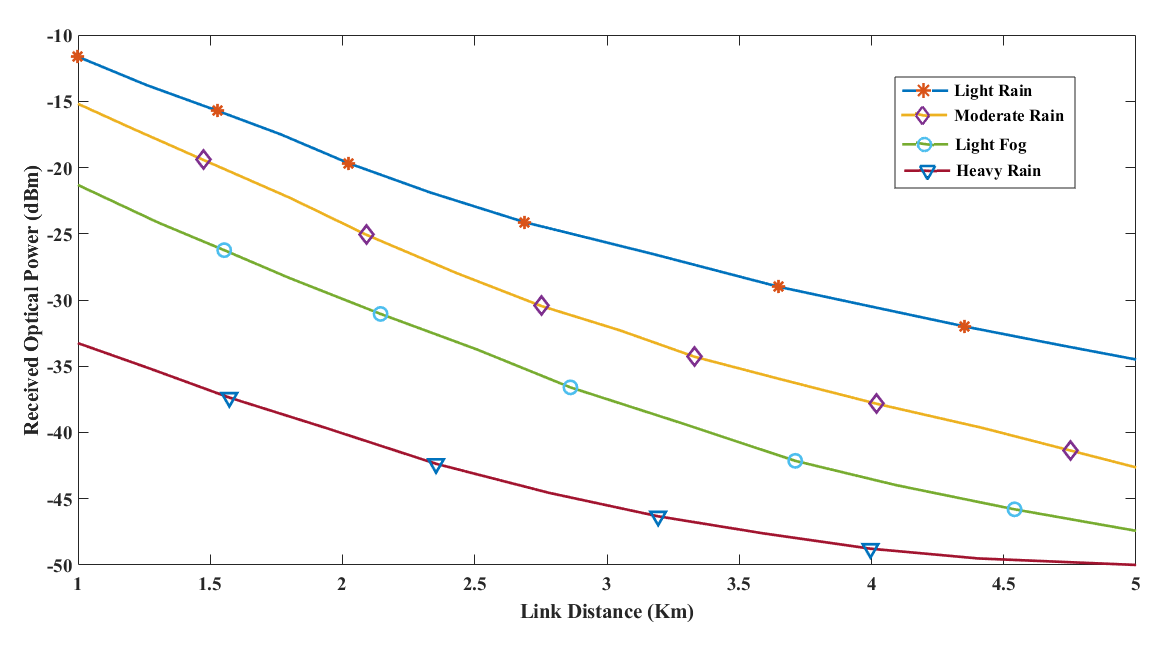} 
	\caption{\centering Received optical power vs. link distance.}\label{Rx-p}	
    \end{figure*}
    
    \begin{figure*}
	\centering
	\includegraphics[width=7in,height=5in,keepaspectratio]{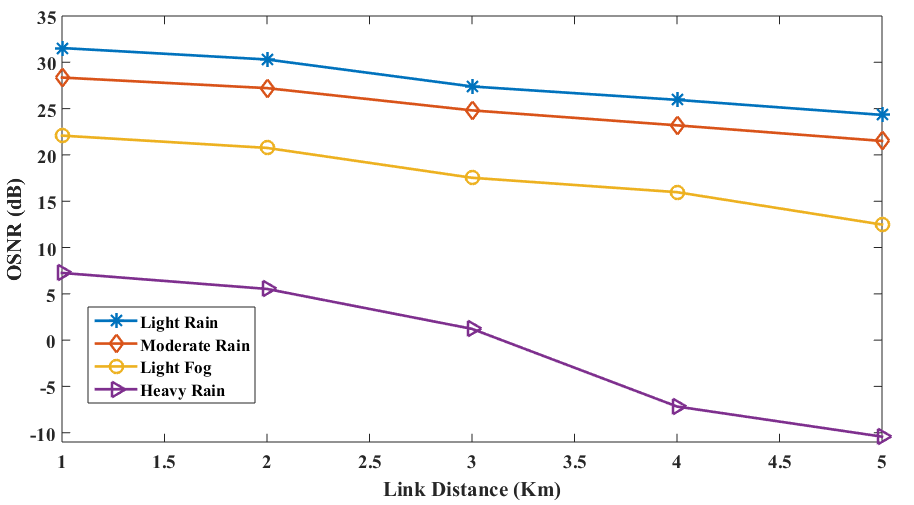} 
	\caption{\centering OSNR vs. link distance.}\label{OSNR}	
    \end{figure*}
    
    \begin{figure*}
	\centering
	\includegraphics[width=7in,height=9in,keepaspectratio]{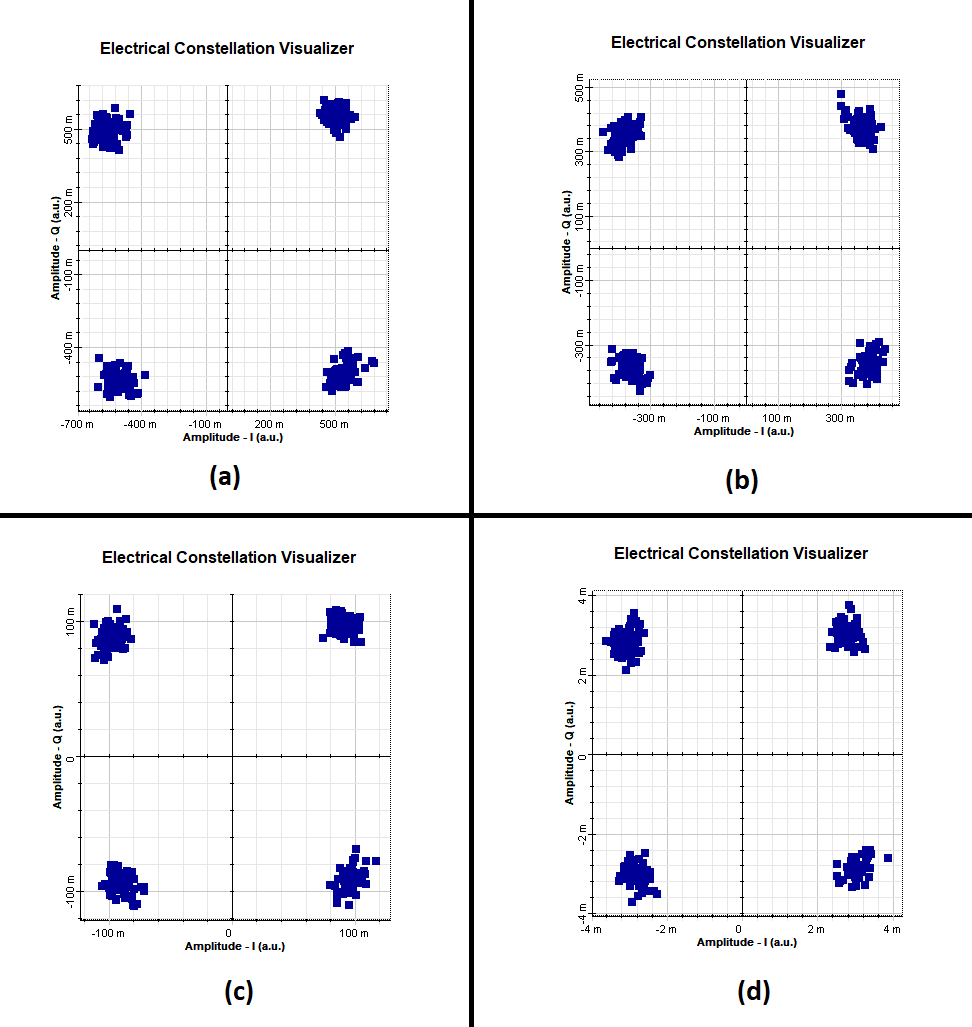} 
	\caption{\centering Constellation diagram (received signal) of our proposed hybrid system at 3 km under: (a) light rain (b) moderate rain (c) light fog (d) heavy rain conditions.}\label{const}	
    \end{figure*}

    \begin{figure*}
	\centering
	\includegraphics[width=7in,height=9in,keepaspectratio]{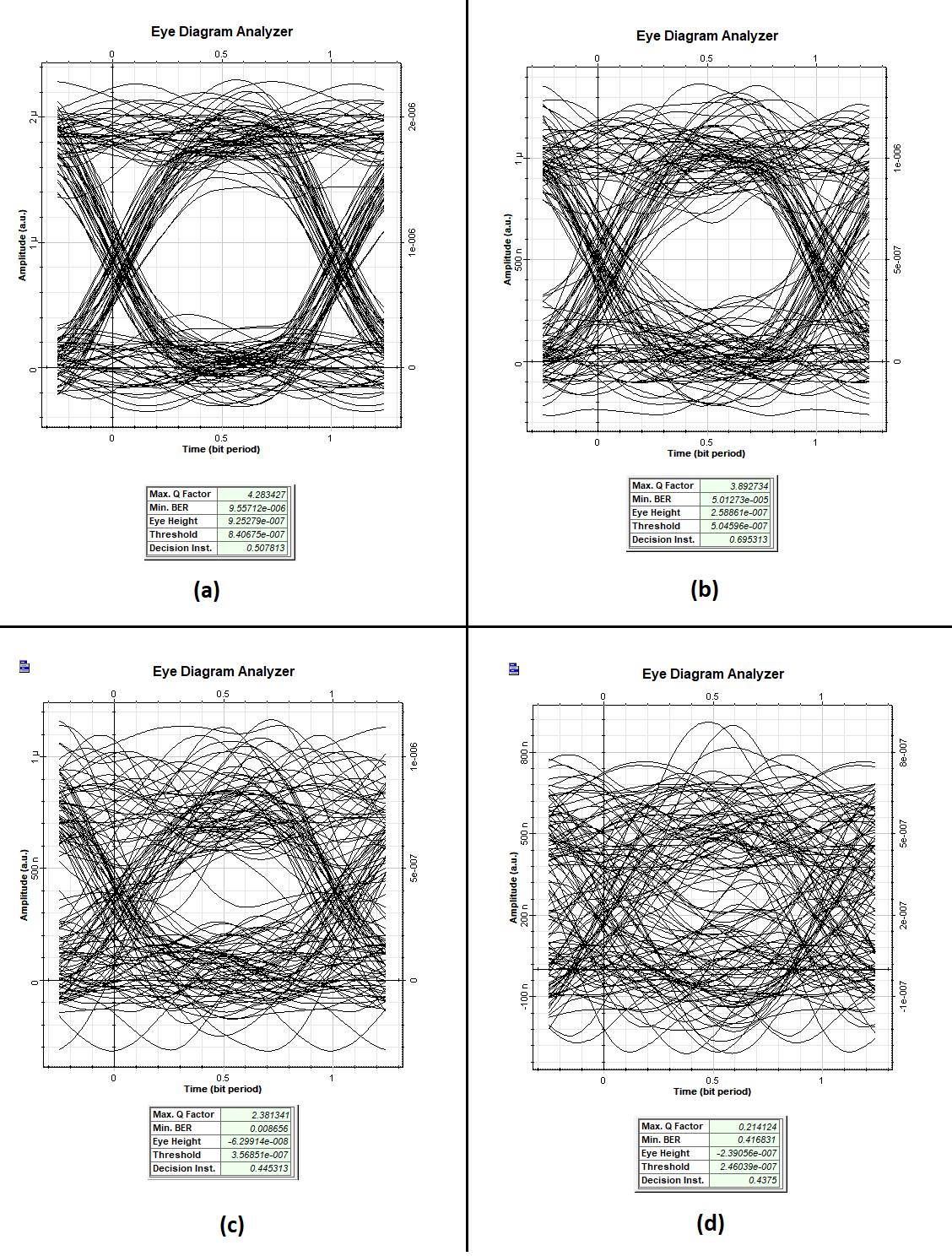} 
	\caption{\centering Eye diagram of our proposed hybrid system at 3 km under: (a) light rain (b) moderate rain (c) light fog (d) heavy rain conditions.}\label{eye}	
    \end{figure*}

 \begin{table}
	\centering
	\caption{Novel Hybrid CPDM-CO-OFDM FSO system parameters and their values.}\label{para}
\begin{tabular}{c c}
			\hline
			\textbf{System Parameter} & \textbf{Value} \\ \hline
			Bit Rate & 200 Gbps\\
		    CW Laser Power & 20 dBm\\ 
		    CW LO Laser Power & 20 dBm\\ 
			CW Laser Linewidth & 10 MHz \\ 
		    CW LO Laser Linewidth & 10 MHz\\
			Operating wavelength & 1550 nm\\
			Link Distance & 1-5 km\\
			Optical amplifier gain & 15 dB\\
			Modulation type & QPSK\\
			OFDM Sub-carrier No. & 128\\
			Used OFDM Sub-carrier No. & 80\\
			No. of Prefix Point & 20\\
			No. of Training Symbol & 10\\
			No. of Pilot Symbol & 6\\
			Gaussian optical filter BW & 500 GHz\\
            Noise Margin & 2 dB\\
			Dark current & 10 nA\\
			Thermal power & $10^{-22}$ W/Hz\\
			Thermal Bandwidth & 10 GHz\\
			PIN Photodiode Responsivity & 0.95 A/W\\
			FSO Transmitter aperture diameter, $d_t$ & 7.5 cm\\
			FSO  Receiver aperture diameter, $d_r$ & 20 cm\\
			Beam Divergence & 2 mrad\\
			Refractive index structure parameter, ${C_n}^2$ & $1.7\times10^{-14}$ $m^{-2/3}$\\
			\hline 
		\end{tabular}
	
\end{table}

	\begin{table}
	\centering
	\caption{Attenuation co-efficient used in the simulation for diverse weather conditions of Bangladesh\cite{7}.}\label{attenuation}
\begin{tabular}{c c}
			\hline
			\textbf{Weather condition} & \textbf{Attenuation (dB/km)} \\ \hline
			Light Rain & 2.97\\ 
			Moderate Rain & 6.55\\ 
			Light Fog & 12.47\\ 
			Heavy Rain & 23.12\\  
			\hline 
		\end{tabular}
	
\end{table}

In the receiver section, the received signal is first filtered using an optical Gaussian filter with 500 GHz bandwidth (BW) then recovered deploying coherent homodyne and finally demodulated to get the output data. The received signal is recovered by a CO receiver shown in Fig \ref{Rx-h} which is split into RCP and LCP components by a CPBS. Each component is then further split into two orthogonal components. Similarly, the local oscillator (LO) signal is split into its orthogonal components using a CPBS and two PBS and is combined with the received signal. The LO is a CW Laser source of the same parameters that area used during transmission. Using eight balanced detectors, each consisting of two PIN photodiodes, the optical signals are converted into electrical signals. Balanced photodetection is used for detecting signal fluctuations, producing high signal-to-noise (SNR) ratio, and canceling laser noise. These electrical signals (In phase \& Quadrature phase) are amplified and send to an OFDM demodulator for demodulation, see Fig \ref{hybrid}. The demodulated signal is further decoded by a QPSK decoder. BER test set, electrical constellation visualizer, polarization analyzer, optical power spectrum analyzer and eye diagram visualizer are used as measurement tools to produce and visualize simulation results.

Fig \ref{OFDM} demonstrates the conceptual OFDM modulation and demodulation techniques. In the transmitter section, serial to parallel conversion is carried out to change the input binary data bits to parallel and later mapped by using an M-ary (QPSK modulator in our model) modulator. Therefore, the binary signal is converted into a digital signal and processed by an IFFT block adding cyclic prefix to it. This prevents sub-carrier overlapping and in return maintains orthogonality avoiding Inter-symbol Interference (ISI). After carrying out parallel to serial conversion, DAC converts the digital signal to analog and sends it through the FSO channel. In the receiver section, the received serial signal is transformed back to digital signal by ADC, converted into parallel removing the cyclic prefix and afterward goes through the FFT operation. Demodulation is executed by using an M-ary (QPSK demodulator in this case) demodulator and the signal is turned back to serial returning the original data.

   \newpage

  \subsection{\textbf{FSO channel Characteristics}}
   The proposed hybrid FSO model is designed and simulated using “OptiSystem 17” optical network simulation software for diverse weather conditions, which provides automatic characterization of the FSO channel. FSO channel comprised of three subsystems: transmitter telescope, free space channel, and receiver telescope. The major goal of FSO system is to obtain a stronger signal which could lead to greater link margin and higher link accessibility. The equation of FSO link (used in OptiSystem) is mathematically defined by \cite{18}:

    \begin{equation}
	    P_{R}=P_{T}\dfrac{{d_{r}}^2}{{(d_{t} + \theta D)}^2}10^{-\alpha D/10}
	\end{equation}
	Geometrical loss which is emerged by spreading the transmitted beam can be expressed by \cite{18}:
	\begin{equation}
	    L_{G}(dB)=20\log\dfrac{{d_{r}}}{{d_{t} + \theta D}}
	\end{equation}
	where, $d_r, d_t, D, \alpha, \theta$ describes receiver, transmitter aperture diameter (m), link distance (km), atmospheric attenuation co-efficient (dB/km) and beam divergence (mrad), respectively.

     \newpage

	\subsection{\textbf{Atmospheric attenuation}}
	 In FSO link, signal quality is severely degraded by the atmospheric attenuation and it is usually influenced by different weather conditions i.e. rain, fog, dust, low clouds, snow, and different combinations of any of them. Table \ref{attenuation} depicts the atmospheric attenuation co-efficients, $\alpha$ (dB/km) for diverse weather conditions of Bangladesh \cite{7}.

 \newpage

\subsection{\textbf{Atmospheric Turbulent Channel}}
Since atmospheric turbulence is a significant factor, during the traveling of an optical signal through the atmosphere, the refractive index is randomly changed along the propagation path. Inhomogeneity in the wind variations, temperature, and pressure give rise to a variation in the refractive index.  Wave front is distorted by the atmospheric turbulence which affects the phase shifts of the propagating optical signals. These distortions give rise to intensity distortions as well, explicated as scintillation \cite{JOC}. Hence, in this article, we have selected the G-G scintillation channel model in OptiSystem as it properly describes the actual channel parameters and the probability density function (pdf) of $I$ is given by \cite{19} and \cite{7},
	
	\begin{equation}
	    P(I)=\dfrac{2(\alpha\beta)^{(\alpha + \beta)/2}}{\Gamma(\alpha)\Gamma(\beta)}I^{\frac{\alpha + \beta}{2} - 1}K_{(\alpha - \beta)}(2\sqrt{\alpha\beta I})\:,I>0
	\end{equation}

    where, $I$ represents the signal intensity, $\Gamma(\cdot)$ is the gamma function, $K_{(\alpha - \beta)}$ is the modified Bessel function of the second kind of order $\alpha - \beta$, $\alpha$ and $\beta$ are the variances of small and large scale turbulent eddies respectively defined by \cite{19} and \cite{7},

    \begin{equation}	
	 \alpha=\bigg[\exp(\dfrac{0.49\delta^2}{(1 +1.11\delta^{12/5})^{7/6}}) -1\bigg]^{-1}
	\end{equation}	
	\begin{equation}
    \beta=\bigg[\exp(\dfrac{0.51\delta^2}{(1 + 0.69\delta^{12/5})^{5/6}}) -1\bigg]^{-1}
	\end{equation}

	Where, $\delta^2$ represents the Rytov Variance and it is defined by,
	
	\begin{equation}
	\delta^2=1.23C_{n}^2k^{7/6}D^{11/6} \nonumber
	\end{equation}

	Here, k=$2\pi/\lambda$ represents the Optical wave number, $\lambda$ is wavelength, D=link distance. Moreover, ${C_n}^2$ changes from $10^{-13}$  $m^{-2/3}$ for strong turbulent conditions to $10^{-17}$  $m^{-2/3}$ for weak turbulent conditions \cite{20}. Refractive index structure parameter ${C_n}^2$ determines the turbulence strength and it clearly depends on the geographical location, altitude, time and wind speed. Using the values of altitude and the wind speed collected from Bangladesh Meteorological Department \cite{7}, the values of ${C_n}^2$ lie between $1.15\times10^{-14}$ to $1.7\times10^{-14}  m^{-2/3}$ for Bangladesh, which is close to the values of ${C_n}^2$ for moderate turbulent conditions. Thus, we have considered and selected $C_{n}^2=1.7\times10^{-14}  m^{-2/3}$ in “OptiSystem 17” for diverse weather conditions of Bangladesh throughout the whole simulation.

\newpage	
	
	\section{\textbf{ Results \& Discussions} }\label{result}
	
	The performance of our proposed novel hybrid CPDM-CO-OFDM FSO system is investigated in this section under diverse weather conditions of Bangladesh using OptiSystem 17.

The fog particles which remain longer in the air severely affect the transmitted optical power and ultimately degrades the FSO link performance more than rain. Fig \ref{Rx-p} elucidates the impact of diverse atmospheric attenuations on the received optical power for the FSO link range up to 5 km. It is found that at 3 km and 4 km link distances, the optical received powers are -25.72 dBm, -31.94 dBm, -37.55 dBm, -45.52 dBm and -30.56 dBm, -37.92 dBm, -43.61 dBm, -48.87 dBm under light rain, moderate rain, light fog, and heavy rain conditions, respectively.

\newpage

The corresponding constellation diagrams (received signal) and eye diagrams at 3 km for different weather conditions are demonstrated in Fig \ref{const} and Fig \ref{eye} repectively. This comparison helps to understand the quality of the received signal in terms of eye height in the eye diagrams and the constellation diagrams of received signals after traveling through the turbulent atmosphere and the degrading effects of attenuation with the deterioration of the weather conditions. Since weather conditions worsen, as expected, the eye height continuously shrinks and the signal eventually distorts which is depicted with the help of constellation diagrams.

Moreover, as the link distance increases, the OSNR performance gradually decreases due to atmospheric attenuation, see Fig \ref{OSNR}. For the heavy rain condition the OSNR falls rapidly and at 5km it becomes close to -10 dB. In case of other weather conditions, the slope of the graph decreases almost linearly. 

\newpage

Similarly, the system is analyzed for BER performance varying link distance from 1-5 km for several atmospheric attenuations, see Table \ref{sum}. It is observed that heavy rain condition produces severe BER performance compared to other atmospheric conditions. At 3km, the values of OSNRs are 27.67, 25.04, 17.81, 1.93 dB and BERs are $9.55\times10^{-06}$, $5.01\times10^{-05}$, $8.65\times10^{-03}$, 0.4168 for light rain, moderate rain, light fog and heavy rain conditions, respectively. Although the system is affected by diverse strong atmospheric conditions, from these analyses it can be estimated that the optimum link range is 3 km to get a better OSNR and BER performance for all weather conditions. Table \ref{sum} summarizes the results of Q factor (linear), OSNR (dB) and BER corresponding to all diverse weather conditions given for hybrid CPDM-CO-OFDM FSO system.

	\begin{table*}
	\centering
	\caption{Q factor(linear), OSNR (dB) and BER of our proposed system under: (a) light rain (b) moderate rain (c) light fog (d) heavy rain -conditions of Bangladesh.}\label{sum}
\begin{tabular}{c c c c c c c}
			\hline
			\textbf{Condition} & \textbf{Link distance} & \textbf{1 km} & \textbf{2 km} & \textbf{3 km} & \textbf{4 km} & \textbf{5 km}\\ \hline
		 
			(a) & Q factor & 6.89 & 5.98 & 4.28 & 3.62 & 3.002\\
			    & OSNR (dB) & 31.61 & 30.29 & 27.67 & 26.20 & 24.81\\
		    	& BER & $2.69\times10^{-12}$ & $1.11\times10^{-09}$ & $9.55\times10^{-06}$ & $1.47\times10^{-04}$ & $1.34\times10^{-03}$\\
		    	
			\hline

			(b) & Q factor & 5.85 & 5.13 & 3.89 & 3.23 & 2.66\\
			    & OSNR (dB) & 28.39 & 27.38 & 25.04 & 23.56 & 21.88\\
		    	& BER & $2.45\times10^{-09}$ & $1.45\times10^{-07}$ & $5.01\times10^{-05}$ & $6.19\times10^{-04}$ & $3.91\times10^{-03}$\\
			\hline

			(c) & Q factor & 4.02 & 3.45 & 2.38 & 1.94 & 1.33\\
			    & OSNR (dB) & 22.26 & 20.97 & 17.81 & 16.48 & 13.02\\
			    & BER & $2.91\times10^{-05}$ & $2.80\times10^{-04}$ & $8.65\times10^{-03}$ & $2.33\times10^{-02}$ & $9.17\times10^{-02}$\\
			\hline
			 
			(d) & Q factor & 0.421 & 0.345 & 0.21 & 0.08 & 0.055\\
			    & OSNR (dB) & 7.83 & 6.16 & 1.93 & -6.34 & -9.33\\
			    & BER & 0.3369 & 0.365 & 0.4168 & 0.4681 & 0.4793\\
			\hline 
		\end{tabular}
	
\end{table*}

 \newpage

\section{\textbf{Conclusion}}\label{conclusion}
The proposed design incorporates a novel hybrid CPDM-CO-OFDM model for the FSO communication system. Hybridizing CPDM with OFDM helps as a very suitable means of maximizing the capacity as well as the SE of the system, and reducing the multipath-fading for the FSO link.  The outcomes of our analysis show that atmospheric attenuation as well as turbulence is one of the major causes that degrade the overall system performance, especially under foggy and heavy rain conditions. But the proposed hybrid model exhibits excellent performance even in the turbulent weather conditions and gives us the estimation of possible limit range needed for the CPDM-CO-OFDM FSO link design. From the investigation, it is observed that after 3 km, the proposed system becomes more vulnerable to weather attenuations. Therefore, the optimum link distance for our proposed system is upto 3 km. At this link range, a comparatively better performance is observed even in the foggy and heavy rain conditions of Bangladesh in terms of OSNR, BER and Q factor. Further research can be conducted to boost the availability of the FSO link by improving the degree of polarization of transmitted optical signal close to perfect and incorporating digital signal processing based coherent receiver to enhance the system performances as well.

\newpage
\section{\textbf{Acknowledgement}}
This research received no funding from any funding sources. The authors declare no conflict of interest and wish to thank the anonymous reviewers for their valuable suggestions.

\end{document}